\documentclass[11pt,paper,twocolumn]{article}
\usepackage[utf8]{inputenc}
\usepackage{amsmath}
\usepackage{amsfonts}
\usepackage{amssymb}
\usepackage[cm]{fullpage}
\usepackage{tikz}
\usepackage{pifont}
\usepackage{wrapfig}
\usepackage{xcolor}
\usepackage{soul}
\usepackage{xfrac}
\definecolor{green}{HTML}{123519}
\definecolor{red}{HTML}{800000}
\definecolor{blue}{HTML}{1C789E}
\definecolor{chocolate}{rgb}{0.82, 0.41, 0.12}

\newcommand{\si}{\sigma}

\newcommand{\varep}{\varepsilon}
\newcommand{\ig}{\includegraphics}
\newcommand{\lw}{\linewidth}
\newcommand{\bi}{\begin{itemize}}
\newcommand{\ei}{\end{itemize}}
\newcommand{\bra}{\langle}
\newcommand{\ket}{\rangle}
\newcommand{\bc}{\begin{center}}
\newcommand{\ec}{\end{center}}

\newcommand{\pref}[1]{(\ref{#1})}

\newcommand{\beq}{\begin{equation}}
\newcommand\encircle[1]{%
  \tikz[baseline=(X.base)] 
    \node (X) [draw, shape=circle, inner sep=0] {\strut #1};}
\newcommand{\eeq}{\end{equation}}

\usepackage{wasysym}
\usepackage{hyperref}
\usepackage{mathtools}
\usepackage{caption}

\DeclarePairedDelimiter\round{\lfloor}{\rceil}
\usepackage{authblk}
\usepackage{titling}
\usepackage{graphicx}	
\usepackage[export]{adjustbox}

\date{}
\author{Victor Herdeiro}
\affil{Department of Mathematics, King’s College, London, United Kingdom.}
\title{\emph{\textbf{Markov chain sampling of the O(n) loop models on the infinite plane}}}
\begin{document}

\twocolumn[{%
 \begin{@twocolumnfalse}
\maketitle

\begin{abstract}

It was recently proposed in \href{https://journals.aps.org/pre/abstract/10.1103/PhysRevE.94.043322}{[Herdeiro\;\&\;Doyon Phys.\,Rev.\,E (2016)]} a numerical method showing a precise sampling of the infinite plane 2d critical Ising model for finite lattice subsections. The present note extends the method to a larger class of models, namely the $O(n)$ loop gas models for $n \in (1,2]$. We  argue that even though the Gibbs measure is non local, it is factorizable on finite subsections when sufficient information on the loops touching the boundaries is stored. Our results attempt to show that provided an efficient Markov chain mixing algorithm and an improved discrete lattice dilation procedure the planar limit of the $O(n)$ models can be numerically studied with efficiency similar to the Ising case. This confirms that scale invariance is the only requirement for the present numerical method to work.

\end{abstract}
\vspace{1cm}

\end{@twocolumnfalse}
}]


\section{Introduction}

Within statistical mechanics, critical models form a very rich and exciting subject. Many models are known to exhibit correlations over infinite distances and divergences of the free energy or its derivatives \cite{mussardo2010statistical}. The renormalization group (RG) and conformal field theory (CFT) have given us an efficient theoretical framework to study and classify such systems \cite{francesco2012conformal}. One of the main success is in explaining how they fall into universality classes where they share universal exponents and identical emerging collective behaviour. In the case of critical lattice spin models, such universal collective behaviour can be the nucleation of arbitrary large ordered domains with random loop geometries as boundaries. These random variables have been described by conformal loop ensembles (CLE) \cite{sheffield, werner} and linked to the CFT algebra \cite{doyon_cle}.

Numerical methods, in particular Markov chain MonteCarlo (MCMC), have been fruitful in investigating critical models by providing numerical checks and new results beyond analytical reach \cite{newman1999monte}. They rely on very few assumptions - essentially ergodicity - thus have the advantage of being generalizable to higher dimensions and many models. It is of great interest to be able to reproduce, through a Markov chain, probability distributions such as the ones discussed in CLE.

Common choices of boundary conditions, such as periodic or generic fixed boundaries, are improper at sampling the infinite volume bulk observables. Indeed, near the critical point the divergence of the correlation length implies that finite size effects and boundary effects will be carried over infinite range.

The work in \cite{herdeiro2016monte} presented a numerical algorithm accurately approximating the finite domain marginal of an infinite plane distribution in the case of the critical Ising model. These marginals were the restriction of the degrees of freedom to a finite sublattice $A$ obtained after integrating out the fluctuating degrees of freedom from infinitely far away to its boundary $\partial A$. Such boundary may be seen as ``holographic'' in the sense that it has the property of encoding all the information in $\mathbb{C} \setminus \overline{A}$. The preliminary work done in \cite{herdeiro2016monte} showed that a chain of discrete lattice dilations, effectively mapping a state inside $A$ to a state on $\partial A$ composed with a rethermalization by lattice updates, reproduced such a holographic boundary, up to effects measured to be negligible. This was a direct consequence of the scale invariance of the Gibbs measure. This algorithm was dubbed UV sampler as it is equivalent to an inverted RG flow approaching the UV fixed point.

It offers the possibility to approximate averages of random variables of the infinite plane with finite support. This includes CFT correlation functions with insertions of operators - at least for the ones writeable as scaling limit of lattice observables - as well as loop variables such as densities, indicator functions, etc. Ref. \cite{herdeiro2016monte} showed success in fitting the CFT data of the critical Ising model: central charge, scaling weights and structure constants.

In this paper, we show how these techniques can be generalized to the $O(n)$ loop gas for $n\in(1,2]$. In this model, locality is lost as the random loops are nonlocal objects. However, we show numerically how keeping appropriate information on the way loops that cross the boundary are connected in $\mathbb{C}\setminus\overline{A}$, allows us to produce a MCMC that reproduces a holographic boundary. For this purpose, we evaluate various scaling dimensions, three-point coupling and four-point functions using our MCMC, and compare with predicted conformal field theory results. This confirms that a UV sampler only requires scale invariance, and not locality.

The plan of the article is as follows: Section \ref{sec:loopGas} explains general aspects of the algorithm, concentrating on the main differences between the Ising and the $O(n)$ model. Section \ref{sec:sectionCorrelators} presents the lattice observables and their connection to the CFT operator algebra, and numerical checks of the two point functions. In section \ref{sec:sectionCorrelators}, numerical checks for dynamical quantities - namely the spin four point function and the structure constant $C_{\varep \si \si}$ - are presented. A conclusion is presented in Section \ref{sec:conclusion}.

\section{From the Ising model to the \boldmath$O(n)$ critical line.}\label{sec:loopGas} In two dimensions the Ising model is probably the simplest lattice model of statistical physics. It consists of a binary variable\footnote{Bold indices will stand for lattice sites variables} $\si_{\bf i}$ taking values in $\{-1,1\}$ at each lattice sites with only first neighbour interactions (for simplicity we consider the situation without external magnetic field). Its Gibbs measure is given by \beq p(\{\si\},\beta) = \frac{1}{Z(\beta)} \, e^{\,\beta\sum \limits_{\bra {\bf i},{\bf j}\ket}\si_{\bf i}\si_{\bf j} }\label{eq:isingModel}\eeq with $\beta$ the unique coupling, the sum running over every pair of neighbour sites and $Z(\beta)$ the partition function normalizing the probabilities.

This model has been extensively studied in the past for being as insightful as it is simple. Its most interesting feature is its second order phase transition, for a coupling value $\beta_c =0.274\,653\ldots$ on the triangular lattice. When sitting on this critical point, the system exhibits scale invariance. The scaling limit can be taken and all the microscopic details of the lattice geometry become completely irrelevant.

We have shown in \cite{herdeiro2016monte} how to successfully sample a finite subsection of the infinite plane of the critical Ising model approximating to any desired level the holographic boundary.

One simple generalization of this model is increasing the dimension of the local variable to a unit vector of dimension $n$, the Ising model being the special case $n=1$. The product $\si_{\bf i} \si_{\bf j}$ gets promoted to $\vec{\si}_{\bf i} \cdot \vec{\si}_{\bf j}$ to conserve rotational invariance, and because of this symmetry it is believed that \cite{henkel2012conformal} the critical point of this model is in the same universality class as the model with partition function \beq Z_{O(n)} = \sum \limits_{\{\si\}} \prod_{\bra {\bf i},{\bf j}\ket} \big(1+x\,\vec{\si}_{\bf i} \cdot \vec{\si}_{\bf j}\big).\label{eq:vectorModel}\eeq Here the first sum represents an integral over the $n-1$ angular variables of each unit vector. For the Ising model, the equivalence between \pref{eq:isingModel} and \pref{eq:vectorModel} is exact even beyond universality. This partition function can be rewritten in loop variables. Indeed, expanding \pref{eq:vectorModel} gives products of the form $$ (x\,\vec{\si}_{\bf i} \cdot \vec{\si}_{\bf j})( x\,\vec{\si}_{\bf k} \cdot\vec{\si}_{\bf l})\ldots$$ where sites  ${\bf i}, {\bf j}, {\bf k}, {\bf l}, \ldots$  are successively, two by two, neighbours. Such terms only survive the angular integration if ${\bf i}, {\bf j}, {\bf k}, {\bf l}, \ldots$ appear an even number of times. One non-vanishing contribution would be
$$ x^6\,(\vec{\si}_{\bf i} \cdot \vec{\si}_{\bf j})\,(\vec{\si}_{\bf j} \cdot\vec{\si}_{\bf k})\,(\vec{\si}_{\bf k} \cdot \vec{\si}_{\bf l})\,(\vec{\si}_{\bf l} \cdot\vec{\si}_{\bf m})\,(\vec{\si}_{\bf m} \cdot \vec{\si}_{\bf n})\,(\vec{\si}_{\bf n} \cdot\vec{\si}_{\bf i}),$$
which integration gives $x^6 n$. It can be pictured as a closed loop which edges join the six lattice sites. Similarly, any contribution to \pref{eq:vectorModel} can be represented as a configuration of closed loops. These loops are defined on the hexagonal lattice dual to the triangular lattice holding the spin variables. The partition function takes the following expression:\beq Z_{O(n)} = \sum \limits_{C \,\in\, {\cal G}} x^{||C||}n^{|C|}\label{eq:loopModel}\ ,\eeq with $x(\beta)=e^{-2\beta}$, ${\cal G}$ being the set of all possible configurations of non intersecting closed loops on the dual lattice, $||\cdot||$ the sum of the lengths of each loop in the configuration and $|\cdot|$ the total number of loops. In other words, $n$ is a loop-weight and $x$ is an edge-weight. Contrary to \pref{eq:vectorModel}, in \pref{eq:loopModel} the parameter $n$ does not have to be interpreted as the dimension of a vector and is not required to be integer any more. From here on it will be assumed to take any values in $[1,2]$, the range of interest for the rest of this work. These models have been proven to go through a second order phase transition on the critical line \cite{nienhuis1982exact}\beq x_c(n) = \frac{1}{\sqrt{2+\sqrt{2-n}}}.\label{eq:criticalCoupling}\eeq

From \pref{eq:vectorModel} to \pref{eq:loopModel}, the vector variables were traded for loop variables. In some sense - from a RG point of view - the vector variables were integrated out, and replaced, for larger scale, by emergent fluctuating degrees of freedom. For MCMC purposes, it is useful to reinterpret \pref{eq:loopModel} as a lattice model of spin variables. It can be looked at as an Ising model with nearest neighbour coupling $\beta = - \frac12 \ln x_c(n)$, and with a non-local contribution to the Hamiltonian: each (closed) boundary between oppositely oriented spin domains will be a `loop' and will add a weight equal to $n$. This has the convenience that some numerical methods from the Ising model can be tuned to accommodate this non-local contribution, Appendix \ref{sec:appendixEvolution} illustrates how Swendsen-Wang (SW) lattice flips can be enhanced to satisfy the equibalance equation of \pref{eq:loopModel}. Further on, Appendix \ref{sec:appendixSampleDetails} details how $O(n>1)$ samples can be generated from Ising samples used in \cite{herdeiro2016monte}. This process is comparable to a physical quench and aims at reducing computation times.

Recall  that the UV sampler in \cite{herdeiro2016monte} manages to construct a chain of holographic boundaries under the two assumptions of scale invariance and locality. This method takes advantage of the scale invariance to apply lattice dilations on $A$ to effectively map states - in a radial ordered CFT sense - from $\partial (\lambda^{-1} A)$ to $\partial A$, where $\lambda>1$ is the dilation parameter. It was proven that a chain composing such dilations would converge to a chain with holographic boundaries on $\partial A$. We suggest the reader to look at \cite[II. A.]{herdeiro2016monte} for details and formalism on this argument.

\begin{figure}
\centering
\ig[width=.89\lw]{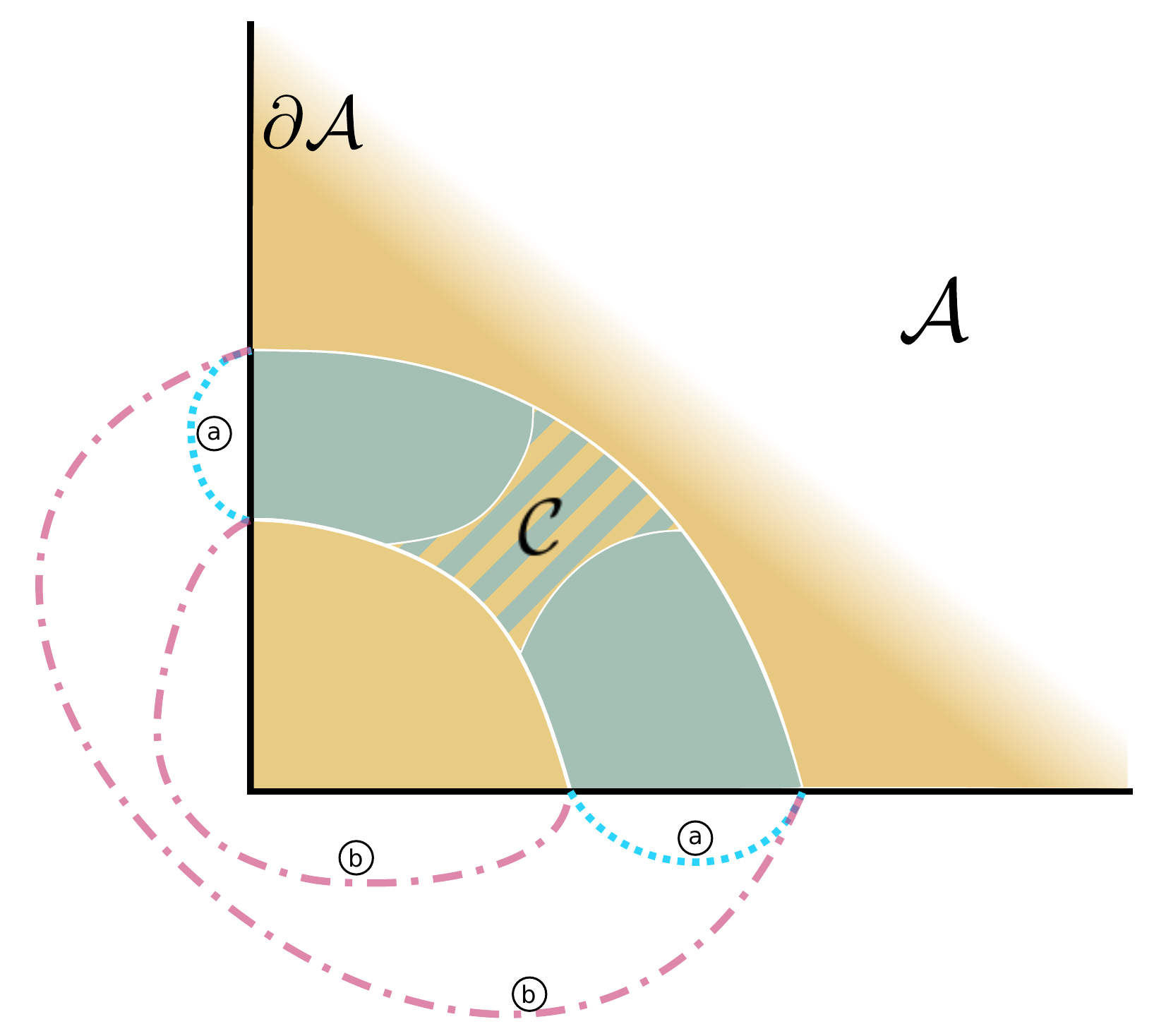}
\captionsetup{singlelinecheck=off}
\caption[foo bar]{\small{We are here looking at a corner of the square domain $A$, which boundaries $\partial A$ are given by the thick straight black lines. The olive and khaki areas are two ordered domains with opposite sign thus the white lines between are arcs of loops entering the partition function. This graph illustrates a case study for a (Wolff)  flip of the $\mathcal C$ virtual cluster from olive to khaki. \encircle{a} and \encircle{b} represent two different, incompatible, connection information beyond the boundary. In the text, we discuss how each information changes the acceptance probability of flipping $\mathcal C$.
}
}
\label{fig:bndFactorization}
\end{figure}

The condition of scale invariance is fulfilled here on the critical line \pref{eq:criticalCoupling}. Compared to the Ising model, the main difference is in the non-locality of the Gibbs measure through its dependence on the total number of loops. When restricted to a finite subsection of the infinite plane it means that the relative weight of two configurations will depend on the loop connections beyond the boundary. In Fig. \ref{fig:bndFactorization}, an example is given. The contribution to the partition function is looked at with two different set of connection information beyond the boundary. If using the information labelled \encircle{a} , the loops are continued by the blue dotted lines. We see that the initial state has 1 loop while the final one has 2. In terms of total number of loops $|\cdot|$ the transition adds one loop and should be accepted with probability $1$ in our chosen algorithm. If using information \encircle{b} instead, continuation by the pink dotted-dashed lines, the transition removes one loop and should only be accepted with probability $n^{-1}$.

The main point is that provided the information of the connections beyond $\partial A$, ratios of partition functions with changes in $A$ are computable.

Compared to our successful chain for the planar critical Ising model \cite{herdeiro2016monte}, a MonteCarlo Markov chain sampling of this measure will need two extra features:

\bi
\item Through each discrete lattice dilation, it needs to track the information on the connections - beyond the border - of the loops touching the border to calculate faithfully the difference $\Delta |\cdot|$ when attempting any flip. Our implementation accounting for this effect is detailed in Appendix \ref{sec:appendixDilation}.

\item Since at each attempted updates we need to compute non-local information, which is the information on the variation of the total loop number, it can be foreseen that the most global updates need to be favoured over the local ones. In this light, lattice updates of the SW algorithm need to be chosen instead of single-cluster or single-spin flips. Our modified SW lattice flips, that take into account loop connections, are introduced in Appendix \ref{sec:appendixEvolution}.
\ei

In this article, we will present numerical evidences that a chain solving the two issues hereabove, in the lattice $O(n)$ models, is able to generate the marginal for the holographic boundaries as it is defined in \cite{herdeiro2016monte}. This will extend the work done in \cite{herdeiro2016monte} to a larger class of non-local models The details of our implementation of a MonteCarlo Markov chain for a lattice dilation operation, a mixing algorithm and the parameters of our MCMC sampling are in appendices \ref{sec:appendixDilation}, \ref{sec:appendixEvolution} and \ref{sec:appendixSampleDetails} respectively. The plan of the article is as follows: Section \ref{sec:sectionCorrelators}  will present the lattice observables and their connection to the CFT operator algebra, and numerical checks of the two point functions and the fitted quantities will be introduced. In section \ref{sec:sectionDynamical}, numerical checks for dynamical quantities - namely the spin four point function and the structure constant $C_{\varep \si \si}$ - will be presented.

%

\section{Lattice operators and correlation functions}
\label{sec:sectionCorrelators}



\begin{figure}
\bc
\ig[width=.9\lw]{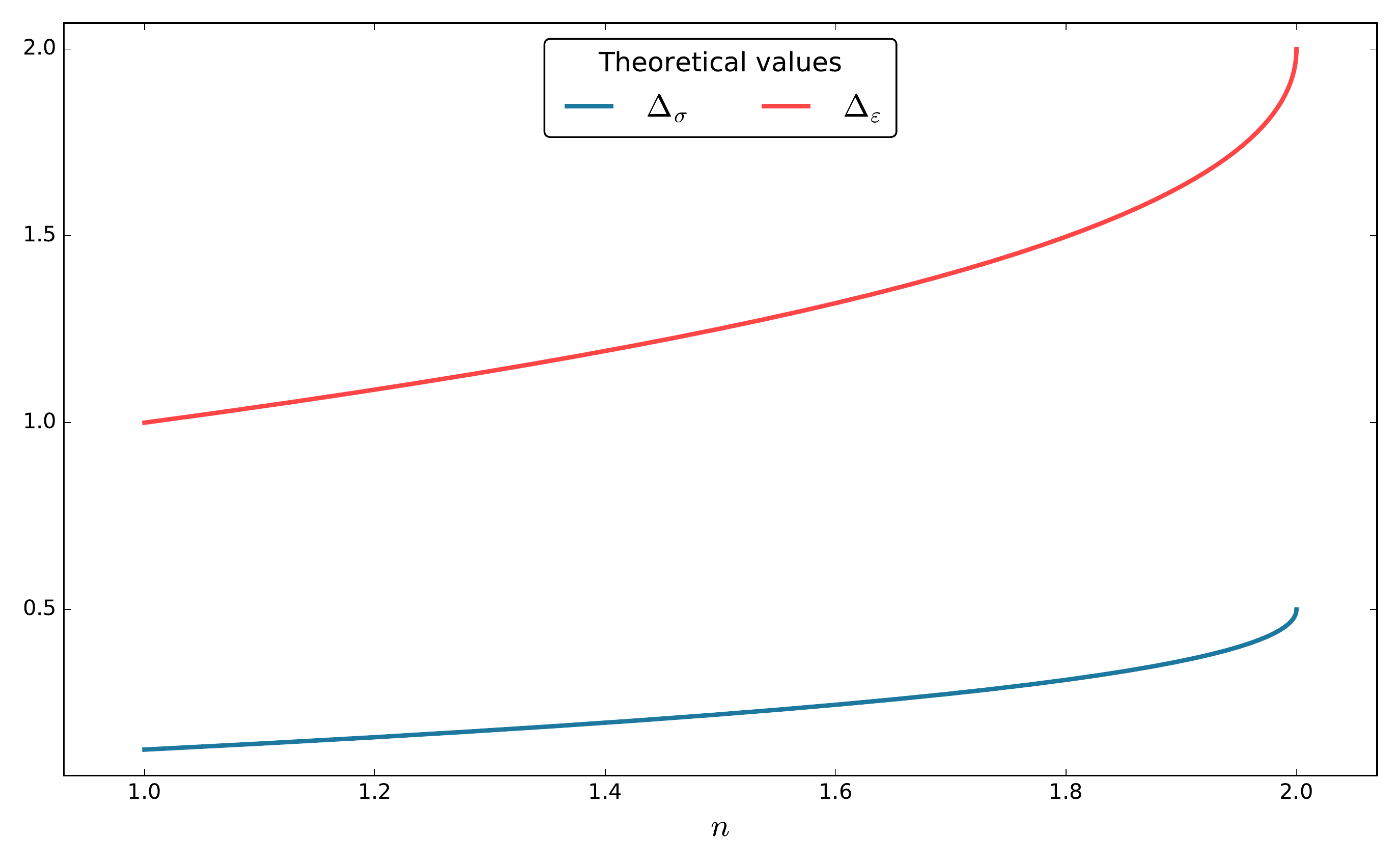}

\caption{\small{The scaling weights of the energy and spin operators have similar profiles, both increase with $n$; this means that correlations decay faster with increasing $n$. Regarding the spin operator, a faster decay with larger $n$ makes sense: the larger $n$ the more energetically favourable it is to create a new loop and thus inducing a flip in the value of $\si_{\bf i}\;\si_{\bf i  + x}$ and thus a decrease of its statistical average.}}
\label{Graph_exponents}
\ec
\end{figure}

\paragraph*{Lattice operators and scaling weights.} The critical line of the $O(n)$ model gives a unique CFT minimal model for each value of $n\in [1,2]$. The scaling exponents for the operators in the Kac table $\{I,\si,\varep\}$ are given by \cite{nienhuis1982exact,nienhuis1984critical}, see Fig. \ref{Graph_exponents}:
\begin{align}\label{CoulombGasExponents}
g(n) &= 2-\frac{\cos ^{-1}\left(-\frac{n}{2}\right)}{\pi }\\
\Delta_{I} &=  0\\
\Delta_{\si} &= \frac{3}{2 g(n)}-1 \label{eq:magneticExponent}\\
\Delta_{\varep}&= \frac{4}{g(n)}-2\end{align}

Each operator here is spinless. Since the $Z_2$ operation $\si_{\bf i }\to -\si_{\bf i}\ \forall {\bf i}$ is a symmetry, they can be classified into even $\{I, \varep\}$ or odd $\{\si\}$ operators. The local lattice variables are the binary lattice variable $\si^L_{\bf i} \in \{-1,1\}$ and the lattice energy density variable $$\varep^L_{\bf i} = \sum \limits_{{\bf j} \in {\cal N}({\bf i})} \si^L_{\bf i} \si^L_{\bf j},$$ with ${\cal N}({\bf i})$ the set of first neighbours of $\bf i$. The $Z_2$ symmetry constrains the CFT operators to be scaling limits of the lattice variables as follows:\begin{align}\si^L_{\bf i} &= a^{\Delta_{\si}} N_{\si}\,\si(a{\bf i}) + \ldots\ ,\\
\varep^L_{\bf i} &= \bra \varep^L \ket I + a^{\Delta_{\varep}} N_{\varep}\,\varep(a{\bf i}) + \ldots\ ,\end{align} with $a$ the lattice spacing -- we will set $a=1$ from here on.

%
%
%
%
%
%

\paragraph*{Correlators.} We use the sampling method detailed in Appendix \ref{sec:appendixSampleDetails}, which is essentially a MCMC with starting point an element from the Ising sampling method in \cite{herdeiro2016monte} sized 2048x2048, mixing to a critical $O(n)$ chain by a ``quench'' done by 8 successive compositions of lattice dilations followed with the appropriate lattice SW flips.

\begin{figure}
\bc
\ig[max width= .9\lw]{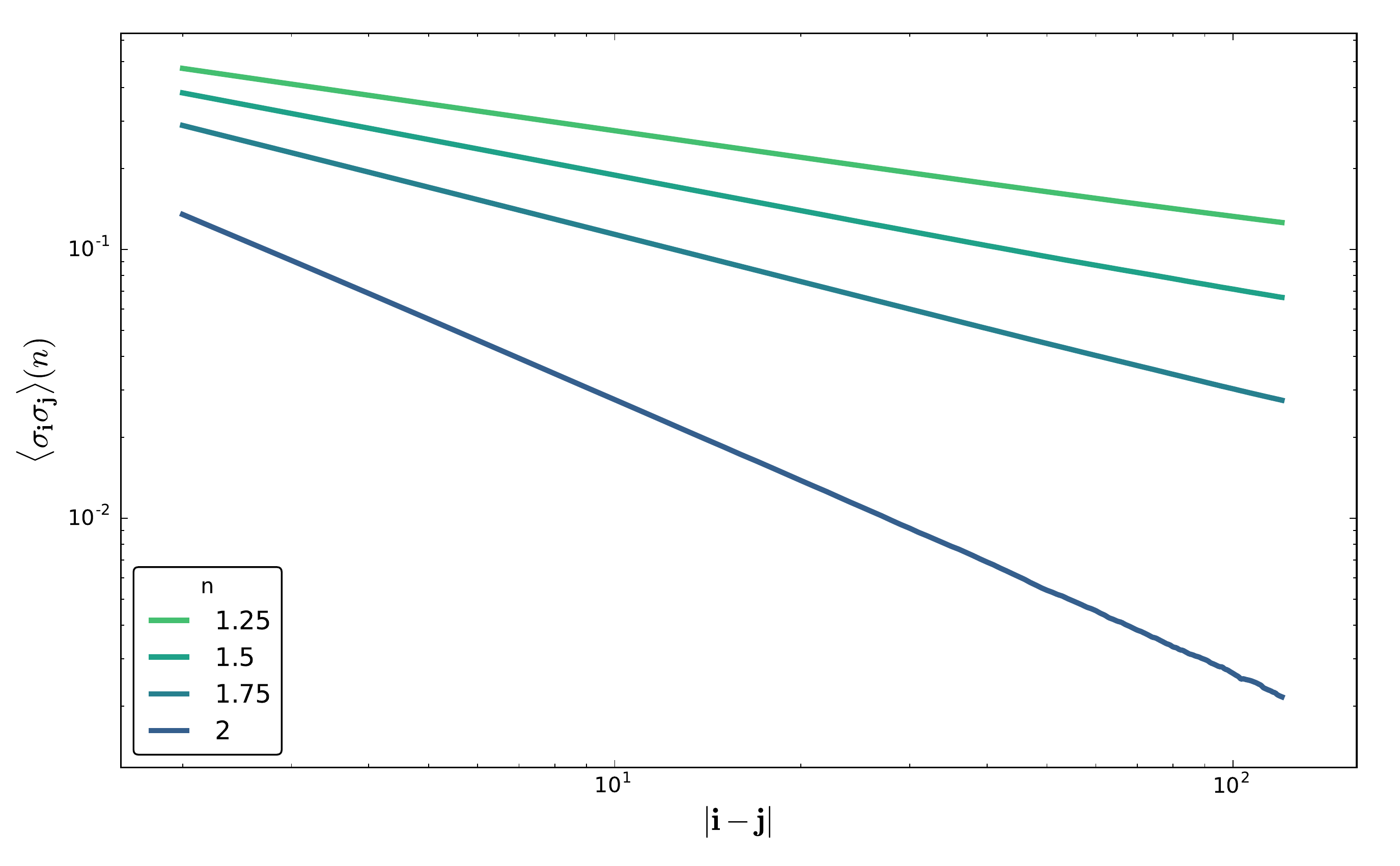}

\caption{\small{Graphs of $\bra \si^L_{\bf i } \si^L_{\bf j} \ket (n)$ for $n=1.25,1.5,1.75,2$ and for separations $|\bf i - j|$ in $[1,120]$. The power law behaviour is manifest. The exponents and offsets fitted are presented in \pref{arrayExponents} and \pref{tab:offsets}. The fits were done by $\chi^2$ minimization on a $x \to ax^b$ template function.}}
\label{SSall}
\ec
\end{figure}

\begin{figure}
\bc
\ig[max width= .9\lw]{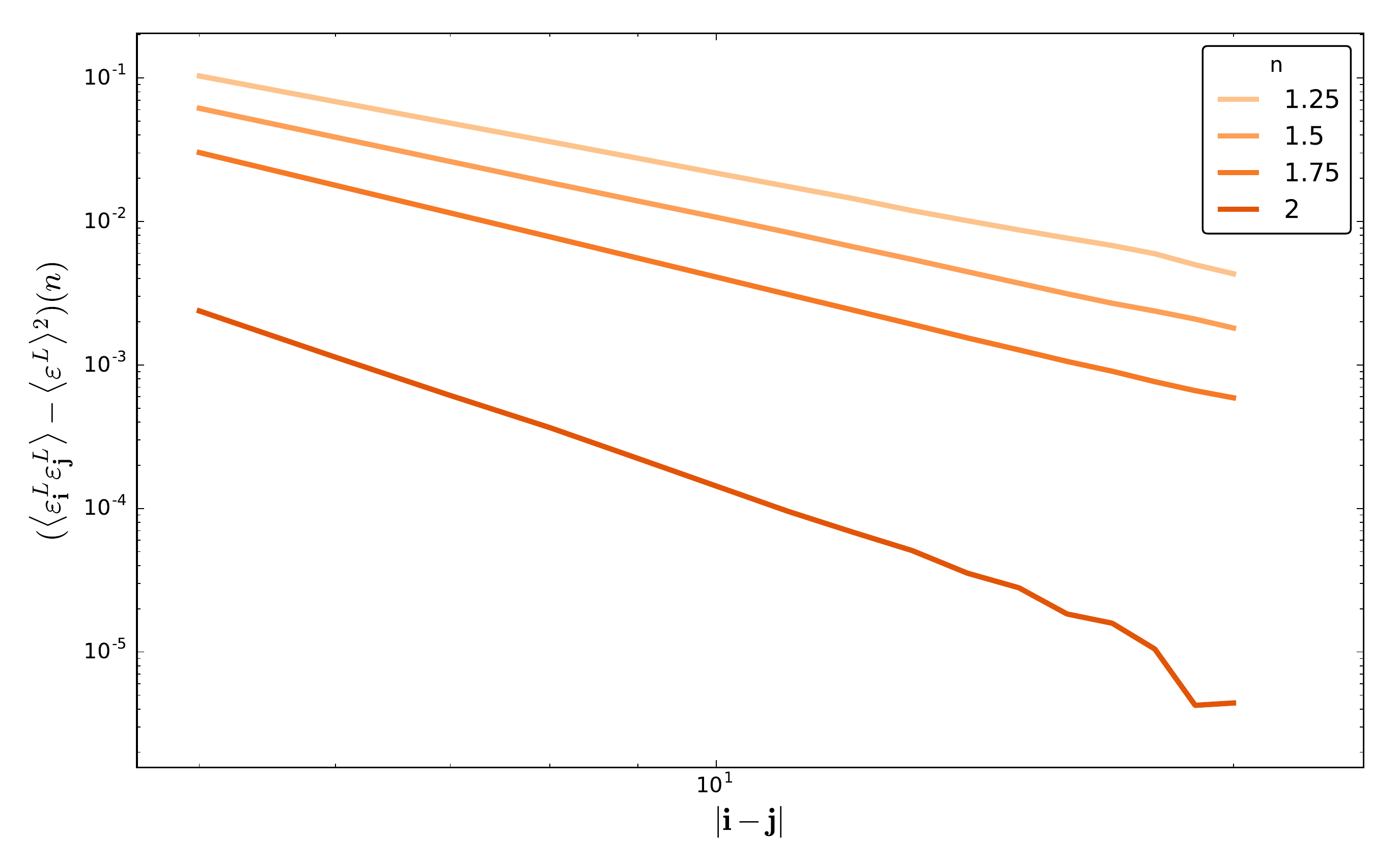}

\caption{\small{Graphs of $\big(\bra \varep^L_{\bf i } \varep^L_{\bf j} \ket  - \bra \varep^L \ket^2 \big)(n)$ for $n=1.25,\,1.5,\,1.75,\,2$ and for separations $|\bf i - j|$ in $[5,20]$. For simplicity, all insertions are made along horizontal direction only. We observe a satisfying power law profile. Only the $n=2$ curve shows some noise in the tail. Fits were performed to extract exponent, amplitude and disconnected part \pref{arrayExponents}, \pref{tab:offsets}. These fits were done by $\chi^2$ minimization on a function $x \to a x^b +c $.}}
\label{EEall}
\ec
\end{figure}

With such chains, lattice two point functions were measured for $\bra \si_{\bf i} \si_{\bf j} \ket$ and $\bra \varep_{\bf i} \varep_{\bf j} \ket$ along horizontal directions. Graphs of the functions are given in Fig. \ref{SSall} and \ref{EEall} respectively. The exhibited power law behaviour is manifest. For $n=1.25, 1.5, 1.75$ and $2$, these allowed to fit the non universal quantities: 

\begin{equation}
\label{tab:offsets}
\begin{array}{l|ccc}
n & N_{\si} & N_{\varep} & \bra \varep^L \ket\\
\hline \hline\\
1.25 & 0.7680\,(5) & 1.941\,(29) & 3.55089\,(4)\\
1.5 & 0.71404\,(6) & 1.8957\,(91) & 3.08255\,(1)\\
1.75 & 0.66011\,(4) & 1.7195\,(96) & 2.557665\,(6)\\
2 &  0.52548\,(48) & 1.205\,(85) &1.540649\,(2)
\end{array}
\end{equation}
These are specific to the triangular lattice.

The following universal observables were also fitted:
\begin{equation}
\label{arrayExponents}
\begin{array}{l|cc}
 n & \Delta_{\si} & \Delta_{\varep} \\
\hline \hline\\
1.25 &  0.1659\,(4) \;\mathbf{\textcolor{chocolate}{0.167225}}& 1.1146\,(43)\;\mathbf{\textcolor{chocolate}{1.1126}}\\
1.5 & 0.21801\,(7) \;\mathbf{\textcolor{chocolate}{0.219459}}& 1.246\,(15) \;\mathbf{\textcolor{chocolate}{1.25189}}\\
1.75 & 0.29136\,(6) \;\mathbf{\textcolor{chocolate}{0.292144}}& 1.4469\,(38)\;\mathbf{\textcolor{chocolate}{1.44572}}\\
2 & 0.4982\,(1) \; \mathbf{\textcolor{chocolate}{0.5}}& 2.0270\,(99)\;\mathbf{\textcolor{chocolate}{2}}
\end{array}
\end{equation}

The bold orange entries are the exact values.

\paragraph*{Geometrical exponents on the bulk.} In \cite{herdeiro2016monte} a method was introduced to estimate numerically the fractal dimensions of clusters in bulk subsections, the method was dubbed bulk finite size scaling (BFSS). These observables have known numerical values \cite{saleur1987fractalexponents,nienhuis1987phase} on the $O(n)$ critical line where the domains become fractal in the scaling limit. On our samples, the following scaling exponent values of the boundary length and domain area were estimated:

%
%
\[
\begin{array}{l|cc}
n & {\rm length} & {\rm mass} \\
\hline \hline \\
1.25 & 1.381\,(3)\;\mathbf{\textcolor{chocolate}{1.389}} & 1.921\,(9)\;\mathbf{\textcolor{chocolate}{1.934}}  \\
1.5 & 1.411\,(4) \;{\mathbf{\textcolor{chocolate}{1.407}}} & 1.938\,(14) \;{\mathbf{\textcolor{chocolate}{1.916}}}  \\
1.75 &  1.427\,(9)\;\mathbf{\textcolor{chocolate}{1.431}} & 1.897\,(3)\; \mathbf{\textcolor{chocolate}{1.903}} \\
2 &  1.502\,(8)\;{\mathbf{\textcolor{chocolate}{1.5}}} &  1.876\,(11)\;{\mathbf{\textcolor{chocolate}{1.875}}}
\end{array} 
\]

The agreement is here satisfying. This is an additional proof that the samples are sitting on the critical line.


\

\section{Dynamical observables in \boldmath$O(1<n\leq2)$}
\label{sec:sectionDynamical}

\paragraph*{Four-point function \boldmath$\bra \si \si \si \si \ket.$}

Correlations functions of 2d CFT in minimal models are constrained to obey some differential equations corresponding to the existence of null states in the Verma modules of the involved operators. The free parameters of these equations are the weights of the fields in the Kac table \cite{belavin1984infinite}. For the $O(n)$ models, the $\bra \si (z_1) \si(z_2) \si(z_3) \si(z_4) \ket$ correlator has been shown to satisfy \cite{gamsa2006correlation}

\begin{figure}
\begin{center}
\ig[width=.9\lw]{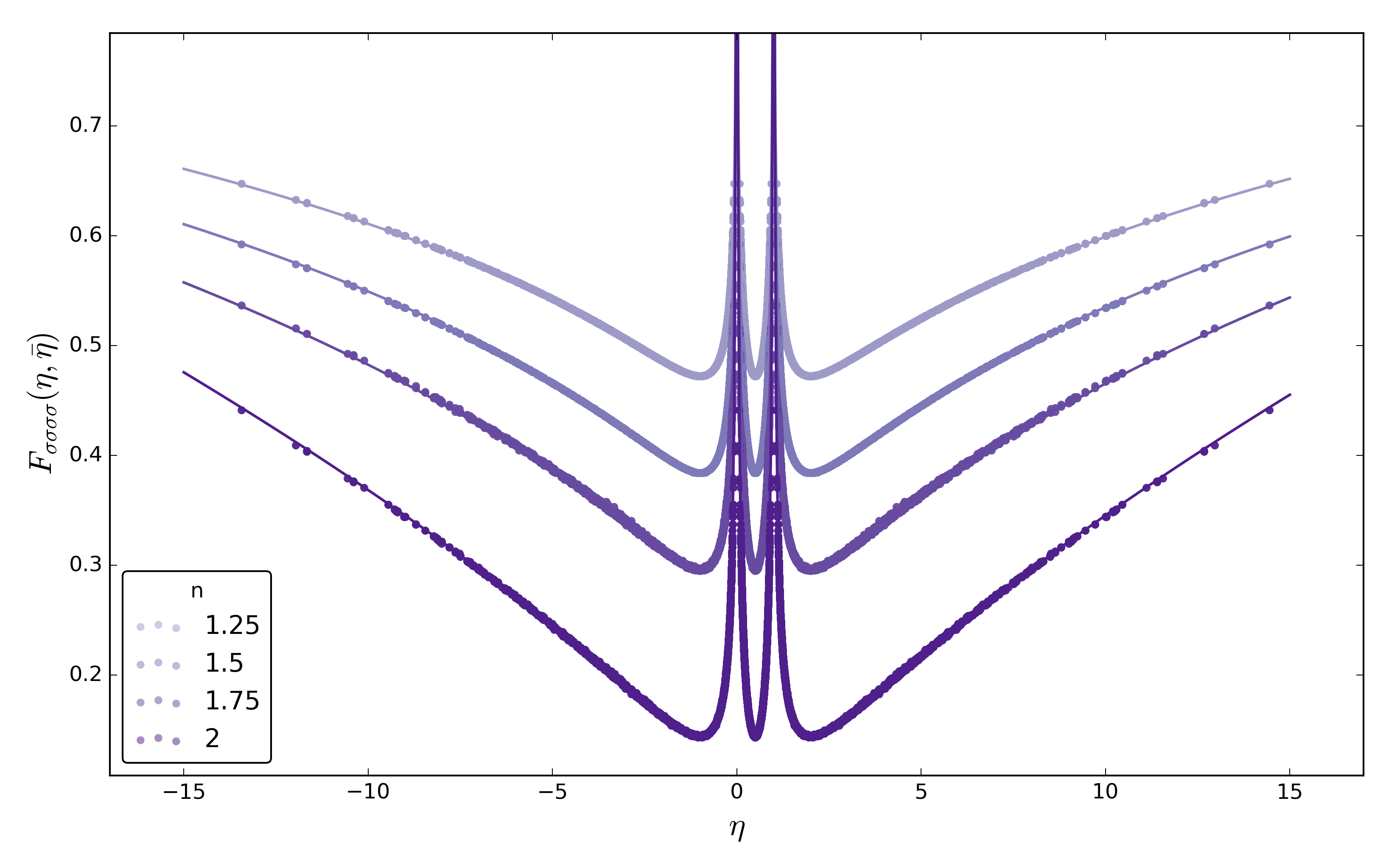}
\end{center}
\caption{\small{Plots of $F_{\si \si \si \si}(\eta = \bar{\eta}, \kappa(n))$ for $n=1.25,1.5,1.75$ and 2. The respective $\chi^2$ values are 0.355, 0.579, 0.671 and 0.614. They share a p-value of 1.}}
\label{SSSSallN}
\end{figure}


\begin{align}\label{eq:FourPointFunction}
\bra \si(z_1) \si(&z_2) \si(z_3) \si(z_4) \ket = \\ &\bigg| \frac{1}{z_{12}z_{13}z_{14}z_{23}z_{24}z_{34}}\bigg|^{\frac{4}{3}h_{\si}} F_{\si\si\si\si}(\eta,\bar{\eta},\kappa)\nonumber
\end{align}
with $$\eta = \frac{z_{12}z_{34}}{z_{13}z_{24}},$$ and $$F_{\si\si\si\si}(\eta,\bar{\eta},\kappa) = \xi(\eta,\bar{\eta},\kappa)\, \big|\eta(1-\eta)\big|^{-\frac{8h_{\si}(\kappa)}{3}}.$$

Here \begin{align*}\xi ( \eta, \bar{\eta},\kappa) &= \Big|{}_2F_1(1-\frac{\kappa}{4},2-\frac{3\kappa}{4};2-\frac{\kappa}{2};\eta)\Big|^2\\ &+ B(\kappa)|\eta(1-\eta)|^{2h} \Big|{}_2F_1(\frac{\kappa}{4},\frac{3\kappa}{4}-1;\frac{\kappa}{2};\eta)\Big|^2,\end{align*}
\begin{align*}B(\kappa) &=\\ &\frac{ \big[\Gamma(1-\frac{\kappa}{4})^2 \Gamma(\frac{\kappa}{4})^2 - \Gamma(2-\frac{\kappa}{2})^2 \Gamma(\frac{\kappa}{2} -1)^2\big] \Gamma(\frac{3\kappa}{4} -1)^2}{ \Gamma(\frac{\kappa}{2})^2 \Gamma(\frac{\kappa}{2}-1)^2 \Gamma(1 - \frac{\kappa}{4})^2}, \end{align*}$$h_{\si}(\kappa) = \frac{3\kappa}{16} - \frac12, \qquad h(\kappa) = \kappa  - 2\ \big(=h_{\varep} (\kappa)\big),$$ $$\kappa(n) = \frac{4}{g(n)},$$ and ${}_2F_1$ the hypergeometric function. This analytical form is a consequence of the existence of a null state at level two in the present CFTs.

Numerically, we aimed at checking the profile of $F_{\si\si\si\si}(\eta, \bar{\eta}, \kappa)$ - this is the purely dynamical part of \pref{eq:FourPointFunction} - restricted on $\eta \in \mathbb{R}$. For simplicity, the insertions were made exclusively along horizontal directions. This was motivated by computational simplicity and also to have a two axis plot. Separations were taken up to 30 lattice sites. Running over our samples, for $n \in \{1.25, 1.5, 1.75, 2\}$, we get Fig. \ref{SSSSallN}. The agreement is striking.

\paragraph*{Three point coupling \boldmath$C_{\varep \si \si}(n).$}

A numerical estimation of the structure constant $C_{\varep \si \si} = \frac12$ was proposed with the Ising UV sampler \cite{herdeiro2016monte}. This was a very interesting check as this constant quantifies the strength of the coupling between the $\si$ and the $\varep$ operators. It encodes a dynamical feature of the model, entirely specific to its universality class. In the same fashion, we wanted to check numerically the value of the same structure constant for the generalized $O(n)$ random loop models. Its analytical expression can be derived from matching the OPE algebra $\si \si = \ldots + C_{\varep \si \si} \varep + \dots$ with an appropriate expansion of $\bra \si \si \si \si \ket$, see Appendix \ref{sec:appendixExpansion}. This gives: $$C_{\varep \si \si}(n) = \sqrt{B(\kappa(n))},$$ with extremal values $C_{\varep \si \si}(n=1) = \frac12$ and $C_{\varep \si \si}(n=2) = \frac{\sqrt{3}}{2}$.

Similarly to the numerical estimation of $C_{\varep \si \si}$ we performed in the Ising model in \cite{herdeiro2016monte}, we repeated an estimation of this structure constant for $n=$1.25, 1.5, 1.75 and 2 by measuring $\bra \varep_{\bf k} \si_{\bf i} \si_{\bf j} \ket$ in a $|{\bf i-j}| \ll |{\bf k -j}|$. This allowed to estimate by fitting on a double power law $x,y \to ax^by^c$:

\beq
\begin{array}{l|cc}
n & C_{\varep \si \si} &{\rm Insertion\,ranges} \\
\hline \hline\\
1.25 &  0.5665\,(108)\quad\mathbf{\textcolor{chocolate}{0.566432}} & [6,11]\quad [65,69]\\
1.5 &  0.6349\,(101)\quad\mathbf{\textcolor{chocolate}{0.633958}} & [6,15] \quad [72,80]\\
1.75 & 0.71075\,(519)\quad \mathbf{\textcolor{chocolate}{0.710603}} & [6,11]\quad [30,41]\\
2 & 0.86560\,(642)\quad \mathbf{\textcolor{chocolate}{0.866025}} & [6,11]\quad [35,67]
\end{array}
\eeq

The ``insertion ranges'' column gives the min and max distance, for $|\bf k-i|$ and $|\bf i-j|$ respectively, taken into account in the fit. The selection was made so as to minimize the $\chi^2$ of the fit, in the sense that the range is picked where the expansion is valid while the signal still dominating over the measurement uncertainty. For each value of $n$, the agreement is satisfying.

\paragraph{Lattice Stress Energy tensor of the \boldmath$O(n)$ loop models.} In \cite{herdeiro2016monte} a lattice representation of the lattice stress energy tensor for the Ising lattice was introduced. This was done by extracting a Fourier mode of spin 2 from the $\si \si$ correlator:$$\si(x) \si(0) = \frac{1}{|x|^{4h_{\si}}}\Big({\mathrm{I} + x^2 \frac{2\,h_{\si}}{c}T(0) + \mathcal{O}(x^3)\Big)} \ + \ \parbox{1.2cm}{energy\\channel.\\terms}$$ Since any primary field will have the identity and its descendants (this includes $T = L_{-2} I$) in its self product OPE, and knowing that $\si$ is still a primary operator in the $O(n>1)$ loop model; this construction should still be valid. This maintains our definition of the lattice stress energy tensor $\cal T$ to be:$${\cal T}_{\bf i} = \sum \limits_{{\bf j} \in \bra {\bf i},\cdot\ket} e^{-2i\theta_{\bf ij}} \si_{\bf i}^L \si_{\bf j}^L.$$


\paragraph*{Lattice Ward identities.} The correlator $\bra T(0) \si(x) \si(y) \ket$ is constrained by the CFT algebra to $$\bra  T(0) \si(x)\si(y) \ket = \frac{h_{\si}}{|{x-y}|^{2\Delta_{\si}}} \frac{(x-y)^2}{x^2y^2}.$$ Such functions entirely defined under the constraint of meromorphicity and the localization of their poles are called Ward identities. In CFT, it includes correlators involving the stress energy tensor, where knowledge of the CFT data is enough to deduce the analytical structure as function of the insertion position of the stress energy tensor.

On the lattice, the Ward identity becomes
\begin{equation}\label{eq:latticeWardIdentity}\bra {\cal T}_{\bf k} \si^L_{\bf i} \si^L_{\bf j} \ket = N_{\cal T} N_{\si}^2 \frac{h_{\si}}{|\bf i - j|^{2\Delta_{\si}}} \frac{(\bf i -j)^2}{(\bf k-i)^2 (\bf k-j)^2} \end{equation}
and if using a prescription inserting the spin operators diametrically opposed to the insertion of the stress tensor, e.g. ${\bf i} - {\bf k} = -({\bf j} - {\bf k})$, \pref{eq:latticeWardIdentity} simplifies to a power law
$$\bra\; {\cal T}_{\bf 0}\, \si^L_{\bf n}\, \si ^L_{-\bf n}\;\ket = N_{\cal T} N_{\si}^2 h_{\si} 2^{2-2\Delta_{\si}} |{\bf n}|^{-2\Delta_{\si} - 2}.$$
Similarly to the definitions $N_{\si}$ and $N_{\varep}$, $N_{\cal T}$ is the scaling factor between the lattice stress energy tensor and its CFT equivalent. 
A fit of the power law's offset and knowledge of $N_{\si}$ from \pref{tab:offsets} offers a numerical estimation of $N_{\cal T}$.

\paragraph*{Numerical central charge.} The central charge is a key parameter of a CFT. Its numerical estimation has been in reach of MCMC methods for instance see \cite{numericalCentralCharge1998}. Following our measurements of lattice correlation functions, it is natural to measure it by looking at the autocorrelations of the lattice stress energy tensor
$$ \bra {\cal T}_{\bf 0} {\cal T}_{\bf n} \ket = \frac{c}{2} \frac{N_{\cal T}^2}{{\bf n}^4}.$$

Fitting this power law's offset and using the previous estimation of $N_{\cal T}$ gives access to a numerical estimation of $c$.  Table \pref{tab:centralCharge} presents the results of the fits of the Ward identity introduced here-above as well as the derived estimations of the central charge for $n=1.25$, 1.5 and 1.75. This computation is similar to the central charge estimation of the Ising model in \cite{herdeiro2016monte}, although the uncertainty of the estimation here could not be brought to a similar low magnitude with a comparable computational effort.

In the estimation of the numerical central charge uncertainty, it was decided to split the error into a systematic error - from the fitting uncertainty of $N_{\cal T}$ - in square brackets, and a measurement error - due to fitting uncertainties - from the power law fit of $\bra {\cal T} {\cal T} \ket$ in round brackets.

The estimations presented in \pref{tab:centralCharge} requiring a significantly longer computational effort than the results introduced previously, we had to resolve to use an improvement of the MCMC presented in Appendix \ref{sec:appendixSampleDetails}. To do so, the colouring SW algorithm was replaced by the algorithm used in \cite{deng2006cluster}; where the authors introduced a Swendsen-Wang algorithm with virtual FK clusters which are not confined inside the loops of the $O(n)$ lattice. In the language introduced in Appendix \ref{sec:appendixSampleDetails}, the direct consequence is a smaller rejection probability when flipping each FK cluster, hence a lower autocorrelation and a more efficient MCMC. Beyond this algorithm upgrade, for the runs giving \pref{tab:centralCharge}, the parameters of the MCMC are identical to the ones given in Appendix \ref{sec:appendixSampleDetails} except for the sample size increased to 1\,000\,000.

For the same computational effort, the $n=2$ chain is dominated by noise and does not allow to run the equivalent fits with decent precision. This data point is postponed.

\twocolumn[{
\begin{@twocolumnfalse}
\centering
\beq
\begin{array}{l|ccc}\label{tab:centralCharge}
n & \bra {\cal T} {\cal T} \ket\ {\rm offset} & N_{\cal T}\ {\rm from}\ \bra {\cal T} \si \si\ket\ {\rm offset} & c \\
\hline \hline\\
1.25 & 0.4837\,(148) & 1.243\,(84) & 0.626\;(19)\,[85] \quad \mathbf{\textcolor{chocolate}{0.6205}}\\
1.5 & 0.5078\,(587) & 1.1773\,(217) & 0.733\;(85)\,[27] \quad \mathbf{\textcolor{chocolate}{0.7418}}\\
1.75 & 0.5235\,(829) & 1.1205\,(452) & 0.834\;(132)\,[67] \quad \mathbf{\textcolor{chocolate}{0.8663}}
\end{array}
\eeq
\end{@twocolumnfalse}
}]

\section{Conclusion}
\label{sec:conclusion}

The results presented here assert of the successful generalization of the recipe introduced for the Ising model in  \cite{herdeiro2016monte} to a class of models highly non-local in the spin variables. This numerical study is not exhaustive and many numerical checks are still missing such as the four point correlators $\bra \si \si \varep \varep \rangle$, $\bra \varep \varep \varep \varep \ket$, the structure constant $C_{\varep \varep \varep}$, or the precise profile of the Ward identities.

Other generalizations are being studied by the author. Extending the recipe to higher dimensions - for instance the 3d Ising model - could offer a new numerical approach to 3d CFT \cite{herdeiro2017ising3d}.

Another generalization would include perturbed CFTs on a lattice with finite correlation length $\xi$ comparable or larger than the linear lattice size $L$. Such numerical simulations usually suffer from strong finite size and boundary effects. Even though the perturbation will break the scale invariance - which is the main ingredient to this numerical recipe taking advantage of lattice dilations - preliminary results show that a tuning of the perturbation coupling after each dilation is enough to construct a Markov chain sampling the bulk marginal of this massive QFT.

\subsection*{Acknowledgements}

The author would like to thank B. Doyon for suggesting this problem as well as his continuous guidance throughout. The author is funded by a Graduate Teaching Assistantship from KCL Department of Mathematics.

\newpage

\appendix

\section{Discrete lattice dilations and tracking loop connections}
\label{sec:appendixDilation}

In ref. \cite{herdeiro2016monte} to take advantage of the scaling invariance and to approach the UV fixed point, we introduced an ``inverse Kadanoff block-spin transformation''. Essentially it is a discrete lattice dilation with parameter $\lambda > 1$, mapping $\lambda^{-1} A \to A$. More precisely it mapped \beq\si_{\bf i} \leftarrow \si_{\round{\lambda^{-1} {\bf i}}}\label{eq:spinDilationMap}\eeq where $\round{\bf x}$ means the closest lattice site to $\bf x$. On a discrete system it is obvious that $A$ holds more information than $\lambda^{-1} A$, implying that \pref{eq:spinDilationMap} cannot be done directly in a one-to-one fashion. A prescription detailed in \cite[II. C. \& Fig. 4]{herdeiro2016monte} forces \pref{eq:spinDilationMap} to be performed one-to-one and fills in the missing information, e.g assigning a spin value on the sites with no preimage, using a heat-bath weighted assignation. It appeared to be the most effective choice for the Ising model. This step takes place between the dilation and the rethermalization.
 
In the $O(n)$ case, the main concern of the discrete dilation procedure needs to be to track the information of the loop connections before/after dilating. The best effort should entirely conserve this information. Heat-bath assignations as we used in \cite{herdeiro2016monte} cannot fulfil this requirement as it implies a non-zero probability of seeing a boundary spin given a value opposite to all its neighbours. Such occurrence would create a loop as an \textit{artefact} of the dilation. Such loop would touch the boundary but would have no antecedent through the dilation and thus no connection information.

Another prescription relaxing the one-to-one requirement on \pref{eq:spinDilationMap} and allowing a one-to-many mapping, e.g. duplication of a spin value by being mapped to all the spins sharing the same dilation preimage, resolves the issue. No \textit{artefact loop} can be created this way.

One necessary improvement to the dilation algorithm is to track the ($\mathbb{C} \setminus A$) connections existing between edges touching the border. On the technical level, the data container we chose for storing such information is a symmetrical map object: a binary tree with a hash table mapping an edge object to another one. This gives a map: \begin{align*}{\rm edge\, X }\to{\rm edge\,Y},\\{\rm edge\, Y }\to{\rm edge\,X}.\end{align*} We call this a connection map object. It makes it easy when tracking a loop by jumping along its edges to jump from an edge touching the border (edge X) to the next one on the same loop and inside $A$ (edge Y).

\begin{figure}
\begin{center}
\ig[width=.82\lw]{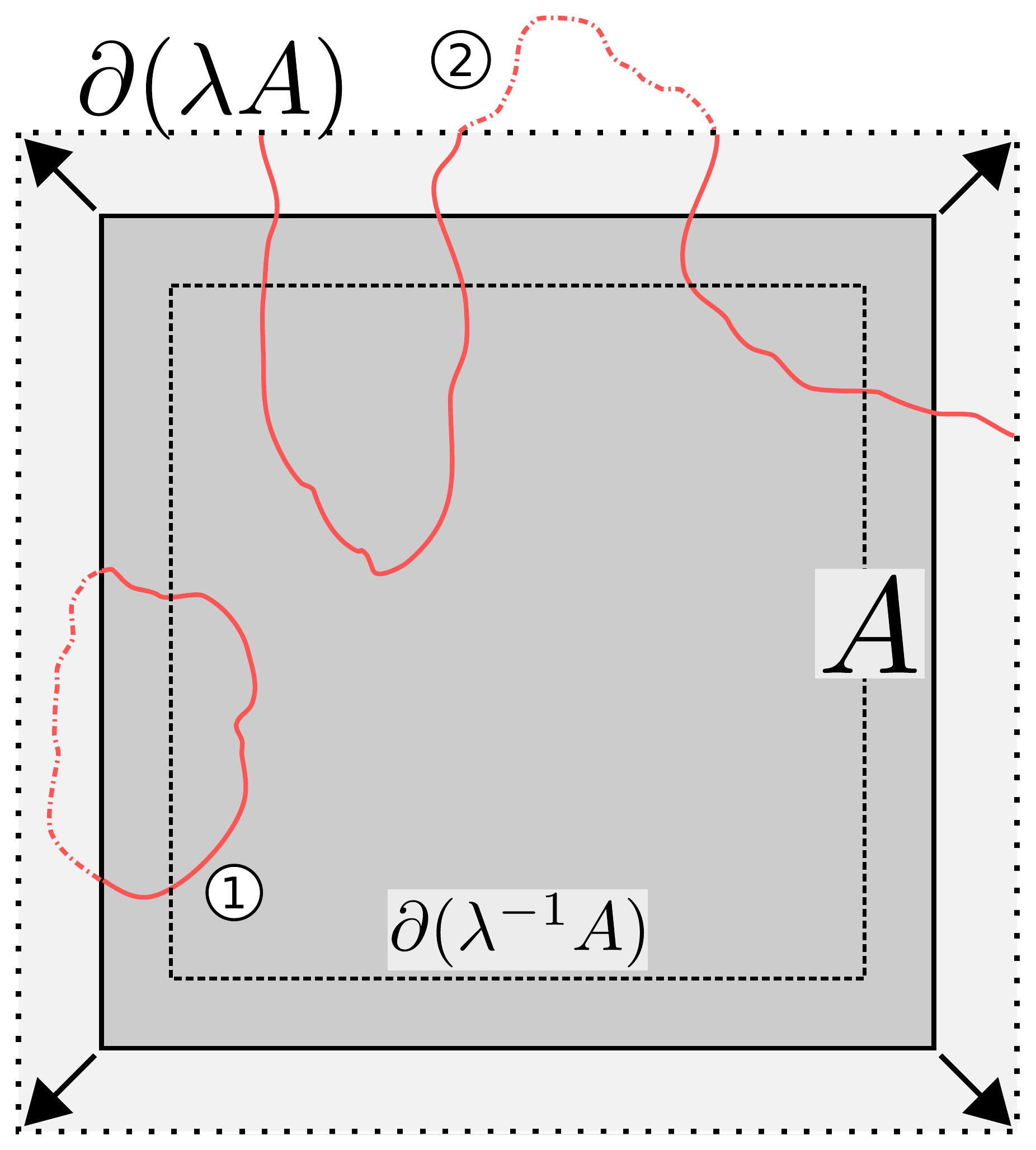}
\captionsetup{singlelinecheck=off}
\caption[foo bar]{\small{Picture of $A$ being dilated to $\lambda A$, $\lambda \approx 1.2$ here. $\partial (\lambda^{-1}A)$, the dashed contour square, gets mapped to $\partial A$, the continuous contour square. Before integrating out $\big(\partial A,  \partial (\lambda A)\big)$ - the light gray area - by cropping it, all information in the loops crossing $\partial A$ needs to be stored in a new connection map object. Cases \encircle{1} and \encircle{2} are described in the text.
}
}
\label{fig:DilationPicture}
\end{center}
\end{figure}

The next step has to be filling this map with the information we get after dilating. It means tracking the information just before cropping the central domain. This is explained by Fig. \ref{fig:DilationPicture}. When updating the boundary connection information post-dilation we see two scenarios here:
\bi
\item case labelled \encircle{1} on the graph: the loop is entirely inside $\lambda A$ and when reducing to $\partial A$ we add to the connection map that the two edges touching $\partial A$ are connected. Connection information is `created' here.
\item case \encircle{2}: this loop touches $\partial( \lambda A)$ thus it is needed to use the information from the pre-dilation connection map object (dotted red line going out of $\partial (\lambda A)$ telling us of a connection between the two pieces of the same loop). This information is still relevant to the connectedness of the loop parts contained inside $\partial A$. Here connection information is updated.
\ei

\section{Swendsen-Wang colouring algorithm}
\label{sec:appendixEvolution}
\subsection*{Evolution algorithm}


As stated before, the motivation in describing the $O(n)$ loop models with spin variables is in the re-usability of the numerical methods known on the Ising model. Here we present how the SW evolution algorithm can be generalized for a $O(n)$ Markov chain. 

In the case of a MCMC on the Ising model, the literature offers many options when choosing the evolution algorithm. We usually distinguish local updates changing the value of a single lattice spin at the time (Metropolis, Glauber, ...) from non-local updates updating one or many clusters at each step (Wolff, Swendsen-Wang, ...). At the critical point, the latter have proven to be very efficient against critical slowdown \cite{wolff1989collective, swendsen1987nonuniversal}. Nonetheless these algorithms are not  straightforwardly generalizable to $O(n)$ loop gas models. The culprit is obviously the non-local contribution of the total number of loops into the Hamiltonian to the flip acceptance probability. This implies directly that any flip, for instance Metropolis or Wolff, will be accepted depending on the variation of the total number of loops. In the pessimistic scenario, calculating this shift in the number of loops requires a computational effort scaling with the lattice area for each attempted flip. This is very ineffective for local updates.

The work done in \cite{ding2007geometric} offers an efficient and simple way of circumventing this obstacle. For $n>1$, we know that the measure \pref{eq:loopModel} favours configurations counting more loops. It can be done using a $n$-dependent freezing of a subset of the loops. Freezing here means that the spins on both sides of each edges of a given loop become non-dynamical, e.g. cannot change value,  with respect to the next spin- or cluster- flip update. Qualitatively this makes the deletion of a loop less likely hence rewards the configurations with more loops. This addon enables any algorithm  - such as Metropolis, Wolff or Swendsen-Wang - to be ``biased'' as to satisfy the equibalance derivable from \pref{eq:loopModel}. This can be proven to happen for a freezing probability of $$p_{\rm freezing} = 1 - \frac{1}{n}.$$The authors of \cite{ding2007geometric} dubbed this prescription ``colouring algorithm'', where colouring a loop means freezing it as defined by the paragraph here above. The name colouring algorithm will be used from here on. Our implementation choice was to couple it to Swendsen-Wang (SW) lattice updates \cite{swendsen1987nonuniversal}.

From the lattice spin variables point of view, the $O(n)$ loop model can be pictured as an - off critical - Ising models with random fluctuating boundaries.


Our lattice flip update procedure follows two steps:
\bi
\item First, we read the loops individually, connecting different segments of a same loop using the connection map information for the loops touching the border. For each loop an independent random choice is made: with a probability $$p_{\rm colouring} = 1-\frac{1}{n}$$ the loop is coloured, meaning that for each of its edges, the spins on both sides will be frozen with respect to the next step.
\item Second, we will run a complete SW lattice flips where only the unfrozen and non boundary sites will be part of virtual FK (\textit{flippable}) clusters. The FK bonding probability is: $$p_{\rm bond} = 1 - x_c(n)$$ with $x_c(n)$ defined in \pref{eq:criticalCoupling}. For each virtual cluster, we will calculate the energy difference $\Delta E$ involving all the edges linking the spins inside the virtual cluster to fixed edges, namely the ones joining them to a border spin or to a frozen spin. Finally, the cluster may be flipped depending on a Glauber acceptance ratio: $$p_{\rm flip} = \frac{1}{1+e^{\Delta E}}.$$


\ei

\subsection*{Numerical Checks}

In \cite{ding2007geometric} the colouring prescription is coupled to Metropolis updates while our implementation choice was to add them on top of SW lattice flips. It seems worth providing a numerical evidence of the correctness of the implementation.

\begin{figure}
\bc
\ig[width=.9\lw]{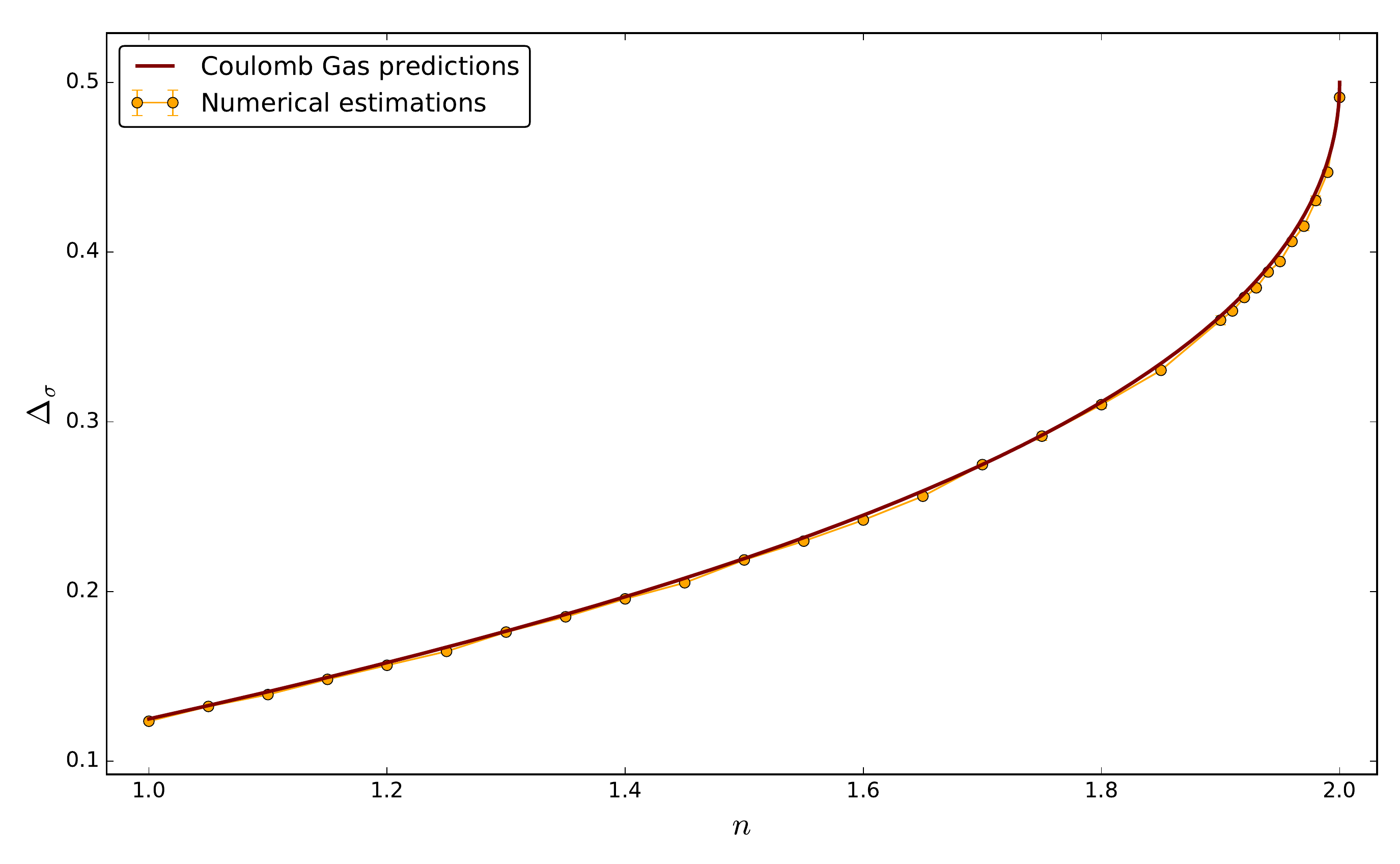}
\ec
\caption{\small{The red line is the graph of the theoretical expression given by \pref{eq:magneticExponent}. The orange points are the fitted scaling exponent values. Each point is the exponent result of a power law fit in the lattice linear size. The uncertainty bars are included but thinner than the data points' width. Here as well we fitted by $\chi^2$ minimization. The agreement is obvious and extremely accurate.}}
\label{fig:FSSmagnetic}
\end{figure}

An easily accessible result is the scaling weight of the spin operator. It is predicted by Coulomb gas methods to give  \pref{eq:magneticExponent}. We fitted the scaling exponent by finite size scaling. When using conformal plus boundaries - equivalent to an external field $h=+\infty$ applied on the boundary spins - RG arguments say that the average magnetization $m = \frac{1}{L^2} \sum \limits_{\bf i} \si_{\bf i}$ should scale as $$m = \frac{1}{L^{\Delta_{\si}}}\ +\ \text{subleading terms}$$ with $L$ the linear lattice size. For $\approx 30$ different values of $n$ in $[1,2]$ and sizes $L$ in $\{32,64,96,128\}$, using coloured SW updates as detailed above and taking 100 000 measurements each separated by 10 updates gave us Fig. \ref{fig:FSSmagnetic}.

This graph is a first supporting evidence of the colouring SW (cSW) evolution algorithm to satisfy the right equibalance equations. On top of that, the chain was fairly effective with very short mixing time and negligible autocorrelations.

For an additional proof, we check if a composition of lattice dilations and rethermalization through cSW updates can create a sample with scale invariance, this succession of steps is more extensively detailed in Appendix \ref{sec:appendixSampleDetails}. This is meant here in the sense that the $\si \si$ correlations will be power law behaved, in the fashion of the results exhibited in \cite{herdeiro2016monte}.

\begin{figure}
\begin{center}
\ig[width=.9\lw]{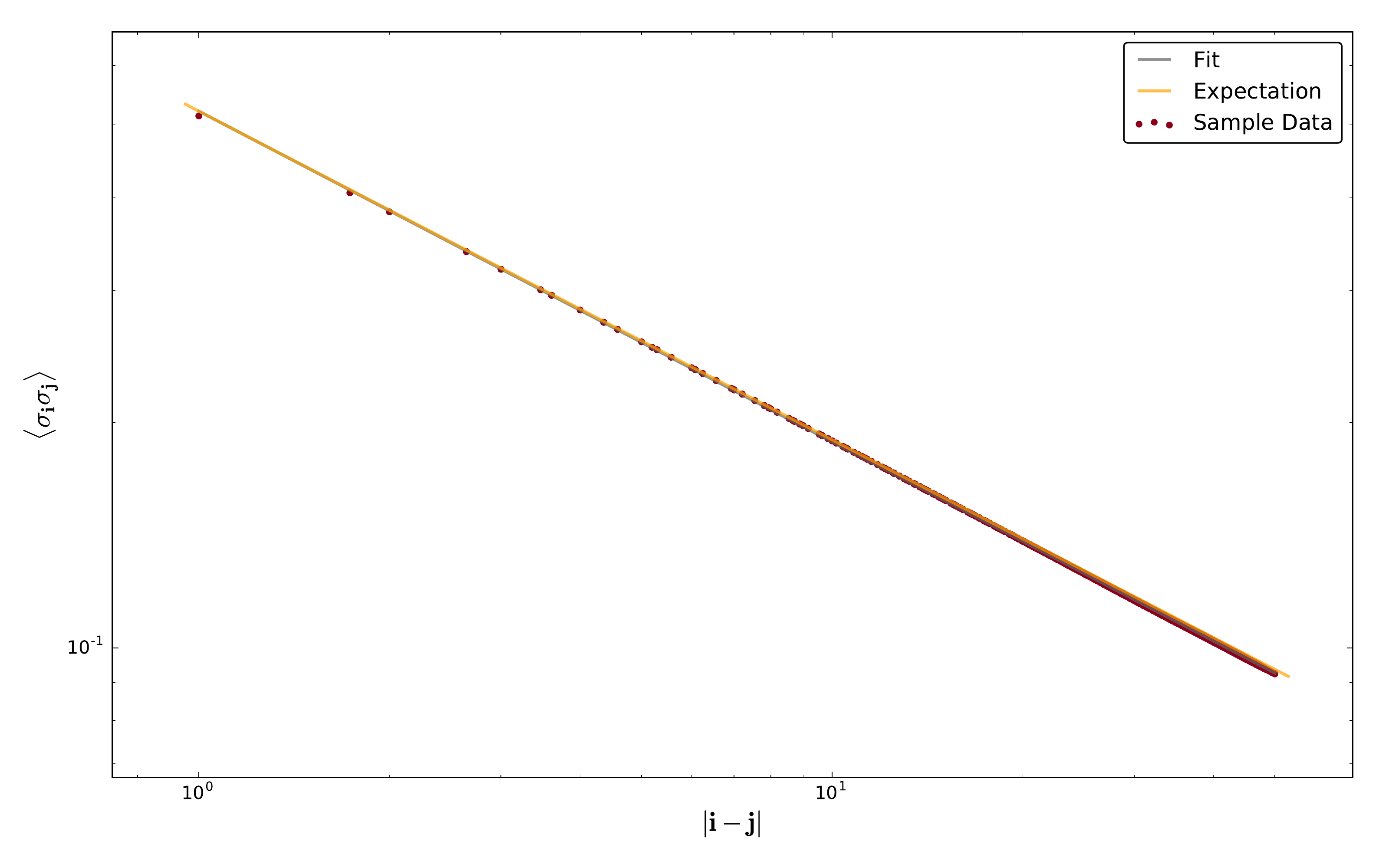}
\end{center}
\caption{\small{On the x-axis, the separation between the two spin insertions. On the y-axis, the value of the average spin correlator. Both axes are log scaled. The power law behaviour seems obvious, the fitted exponent is $2\Delta_{\si} (1.5) = 0.44180\,(5)$ to be compared to the theoretical expectation of $0.43892\dots$ The fit $\chi^2$ is $\approx 10^{-6}$, a very low value.}}
\label{fig:GraphSSplanar}
\end{figure}

For $n=1.5$, a sample of $\sim 1\,500$ lattices sized 2048x2048 is generated. With starting point a sample of planar critical Ising lattices of the same size which is put through a cycle of 10 dilations - $\lambda = 1.2$ - followed by $\approx 50$ coloured SW lattice flips. Appendix \ref{sec:appendixSampleDetails} will present numerical proof that ``quenching'' an Ising sublattice is an efficient way of generating $O(n)$ samples. The correlators were checked with insertions at least 500 lattice units from the boundaries and with separation distances between 1 and 50 to give Fig. \ref{fig:GraphSSplanar}.

These are are first two proofs of the efficiency of our dilation implementation and of the proper scale invariance of our sample. The next appendix presents a deeper analysis of the mixing of the chain as well as a quantitative study of the \textit{bulkiness} of its samples.

\section{Sample parameters}
\label{sec:appendixSampleDetails}

Using the two ingredients above, we present here the details of the MCMC used for all the measurements disclosed in this paper. Ref. \cite{herdeiro2016monte} presented a chain which end products were subsections of the planar critical Ising model on the triangular lattice. Its starting point was the critical Ising model on the torus. For the $O(n)$ we decided to use as starting point these critical Ising samples, since these had been stored and were readily available. The idea is that a fixed number of cycles of lattice dilations followed by coloured SW rethermalization steps, tuned to desired $n$ value, should move the samples by a ``quench'' along the $O(n)$ critical line. In the following, this MCMC will be showed to mix at this precise value of $n$.

\begin{figure}
\begin{center}
\ig[width=.95\lw]{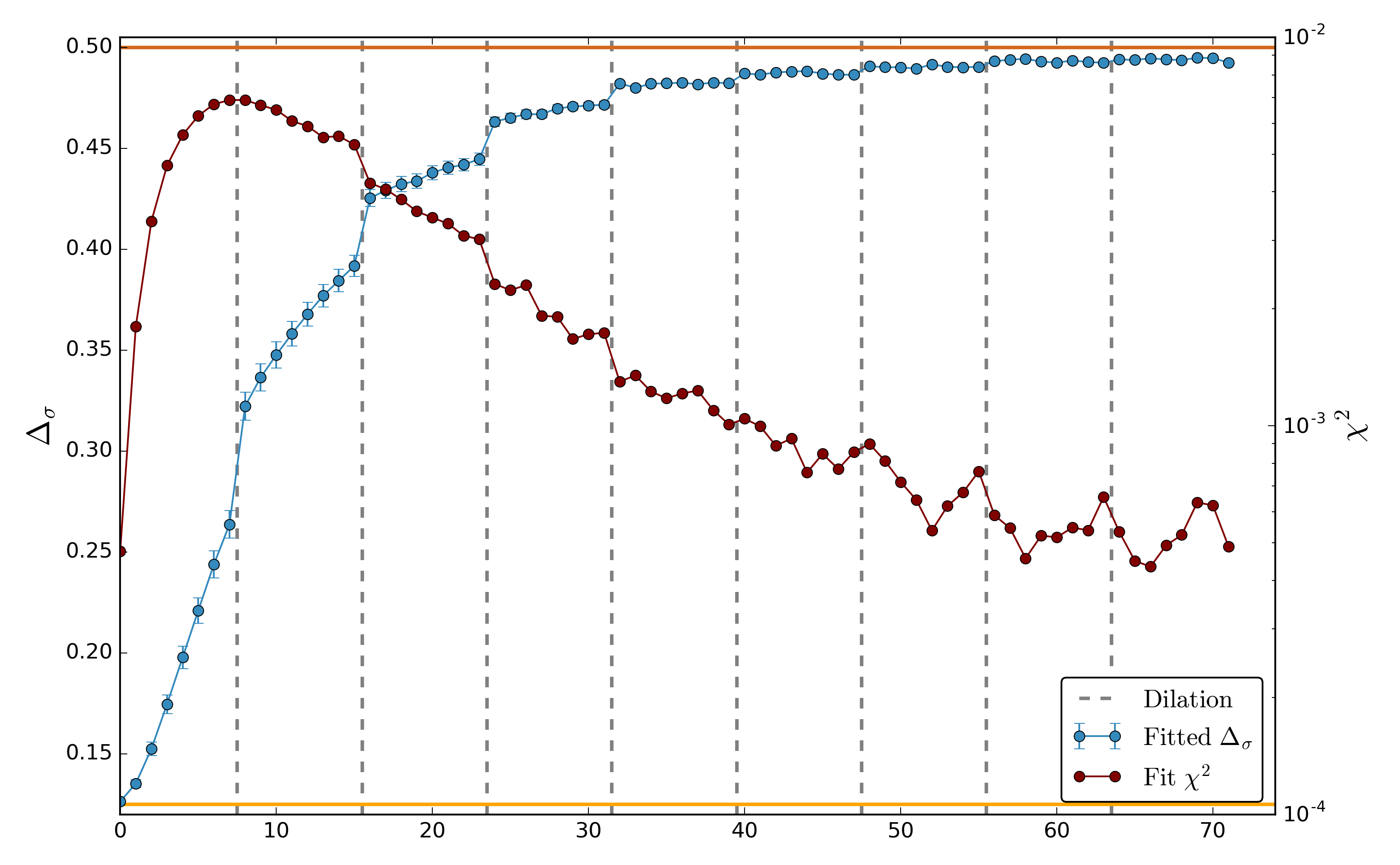}
\end{center}
\caption{\small{\textit{Monitoring} the mixing of a $O(n=2)$ MCMC. Along the x-axis, the chronology of the evolution steps. The vertical dashed lines show the occurrences of a lattice dilation. The y-axis is duplicated: the blue curve shows the value of the fitted $\Delta_{\si}(t)$ with axis markers on the left side; whereas the red curve shows the fits $\chi^2$ with units on the right, the latter is log scaled. The two horizontal lines show the value of $\Delta_{\si}$ for the Ising and the $O(2)$ model, orange and chocolate respectively.}}
\label{Graph_ChainTherma}
\end{figure}

In the case of a MCMC to generate samples of $O(n=2)$ we propose the following timeline:
\bi

\item It starts with a subsection of the triangular critical Ising model, sized 2048x2048.
\item It is followed by $N_{SW} = 80$ cSW lattice flips, tuned to $n=2$ (with fixed boundaries).
\item It will undergo a succession of 8 discrete lattice dilations with parameters $\lambda$ taking values 1.2, 1.2, 1.2, 1.15, 1.1, 1.06, 1.04 and finally 1.02. The motivation is to use large dilations at the beginning to send far away the initial boundaries marginally distributed according to the Ising Gibbs distribution, but they also leave a large number of duplicates on the border ($\sim 17\%$), hence the chain finishes with diminishing values of $\lambda$ to dampen this excess of correlations on and near the boundary.
\item Inbetween each of these dilations, $N_{SW}$ cSW updates are performed to rethermalize the sample.
\item To present a proof that the chain does mix at the $O(n=2)$ critical point, we monitored the $\bra \si_{\bf i} \si_{\bf i+n} \ket$ correlations. The prescription was to take ${\bf i}$ and ${\bf i +n}$ in a central domain, at least 50 lattice sites away from the border, and ${\bf n}$ along horizontal direction with $| {\bf n} | \in [1,50]$ for computational efficiency. One measurement was taken every 10 cSW flips. This was repeated over $\sim 1\,000$ runs and averaged to give a $\bra \si_{\bf i} \si_{\bf i +n} \ket (t)$ at the $t$ step of the mixing process. At each value of $t$ we get a graph of the correlator, $\bra \si_{\bf i} \si_{\bf i+n} \ket$ as a function of $| {\bf n}|$, on which a power law fit is performed to extract an effective value of the scaling weight and an uncertainty on the fit. Here, we use the uncertainty on the fitted scaling weight, linked to the $\chi^2$ uncertainty of the fit.
\item The graph of these monitored quantities along the timeline of the chain is given in Fig. \ref{Graph_ChainTherma}. The graph shows that the fitted exponents indeed travel from the value it takes in the critical Ising model to its value in the $O(2)$ model. The monitoring shows that the dilations and the rethermalization steps work hand in hand: each dilation brings a jump towards the $O(2)$ endpoint. The uncertainty line shows an increase during the transitional period where the correlations have no motivation to be power law behaved. After mixing its magnitude is similar to what it was on the Ising sample, this hints that the end product is of similar quality or \textit{bulkiness} as the chain generating the Ising samples.

\item For each value of $n$ in $\{1.25,1.5,1.75,2\}$, the numerical estimations of the present paper were ran over a sample of size $\sim 50\ 000$. Each sample element was independently generated by a chain just like the one described above.

\ei


\section{$C_{\varep \si \si}$ from $\bra \si \si \si \si \ket$}
\label{sec:appendixExpansion}

In the limit where $z_1 \to z_2$, we have the known OPE:$$ \si(z_1) \si(z_2) = \frac{1}{|z_{12}|^{4h_{\si}}}\ \bigg( 1 + C_{\varep \si \si } \,\varep(z_2)|z_{12}|^{\Delta_{\varepsilon}} + \dots\bigg).$$

Taking the same limit in \pref{eq:FourPointFunction}, an OPE on the left hand side gives: \begin{align*}\bra \si(z_1) \si(&z_2) \si(z_3) \si(z_4) \ket \\ &= \big\bra \frac{1}{|z_{12}|^{4 h_{\si}}} \Big( 1 + C_{\varep \si \si} \varep(z_2) |z_{12}|^{2h_{\varep}} + \dots\Big) \si(z_3)\si(z_4)\big\ket \\&= \frac{1}{|z_{12}z_{34}|^{4 h_{\si}}}\bigg( 1 + C_{\varep\si\si}^2 \bigg| \frac{z_{12}z_{34}}{z_{23}z_{24}}\bigg|^{2h_{\varep}} + \dots\bigg).\end{align*}

For simplicity we set:
\begin{eqnarray*}
z_1 = 0,\quad z_2 \to 0,\quad z_3 = 1,\quad z_4 = 2,
\end{eqnarray*}
giving us:$$\bra \si(z_1) \si(z_2) \si(z_3) \si(z_4) \ket = \frac{1}{|z_2|^{4h_{\si}}} \bigg( 1+ C_{\varep \si \si}^2 \bigg| \frac{z_2}{2}\bigg|^{2h_{\varep}} + \dots \bigg).$$Expanding the right hand side of \pref{eq:FourPointFunction} with $$\eta = \frac{z_2}{2} + O(z_2^2)$$ gives: 
$$\frac{1}{|z_2|^{4h_{\si}}}\bigg(\ldots + B(\kappa)\bigg|\frac{z_2}{2}\bigg|^{2 h_{\varep}} +\dots\bigg).$$The coefficients in front of $\big|\frac{z_2}{2}\big|^{2h_{\varep}}$ read: $$C_{\varep \si\si}^2(n) = B(\kappa(n)).$$

\newpage

\bibliography{merged_bibliography.bib}{}
\bibliographystyle{unsrt}
\end{document}